# Efficient Approximation Algorithms for Spanning Centrality

[Technical Report]

Shiqi Zhang
National University of Singapore
Singapore
Southern University of Science and Technology
China
s-zhang@comp.nus.edu.sg

Renchi Yang[*]
Hong Kong Baptist University
China
renchi@hkbu.edu.hk

Jing Tang
The Hong Kong University of Science and Technology (Guangzhou)
China
The Hong Kong University of Science and Technology
China
jingtang@ust.hk

Xiaokui Xiao[†]
National University of Singapore
Singapore
xkxiao@nus.edu.sg

Bo Tang[†]
Southern University of Science and Technology
China
tangb3@sustech.edu.cn

## ABSTRACT

Given a graph $\mathcal{G}$, the *spanning centrality* (SC) of an edge $e$ measures the importance of $e$ for $\mathcal{G}$ to be connected. In practice, SC has seen extensive applications in computational biology, electrical networks, and combinatorial optimization. However, it is highly challenging to compute the SC of all edges (AESC) on large graphs. Existing techniques fail to deal with such graphs, as they either suffer from expensive matrix operations or require sampling numerous long random walks. To circumvent these issues, this paper proposes TGT and its enhanced version TGT+, two algorithms for AESC computation that offers rigorous theoretical approximation guarantees. In particular, TGT remedies the deficiencies of previous solutions by conducting deterministic graph traversals with carefully-crafted truncated lengths. TGT+ further advances TGT in terms of both empirical efficiency and asymptotic performance while retaining result quality, based on the combination of TGT with random walks and several additional heuristic optimizations. We experimentally evaluate TGT+ against recent competitors for AESC using a variety of real datasets. The experimental outcomes authenticate that TGT+ outperforms state of the arts often by over one order of magnitude speedup without degrading the accuracy.

## CCS CONCEPTS

• **Mathematics of computing** → **Approximation algorithms**; *Markov-chain Monte Carlo methods*; **Graph algorithms**.

## KEYWORDS
spanning centrality, graph traversal, random walk, eigenvector

## 1 INTRODUCTION

Edge centrality is a graph-theoretic notion measuring the importance of each edge in the graph, which plays a vital role in analyzing social, sensor, and transportation networks [5, 11, 32, 37]. As pinpointed by Mavroforakis et al. [29], compared to the classic edge betweenness [9] based on shortest paths, *spanning centrality* (SC) [40] is a more ideal centrality for edges as it accommodates the information from longer paths. In particular, given a connected undirected graph $\mathcal{G}$, the SC $s(e)$ of an edge $e$ is defined as the fraction of spanning trees of $\mathcal{G}$ (a tree-structure subgraph of $\mathcal{G}$ including all the nodes) that contains $e$. In simpler terms, the SC $s(e)$ measures how crucial the edge $e$ is for $\mathcal{G}$ to remain connected, and hence, can be used to identify vulnerable edges in $\mathcal{G}$. Such a definition renders SC useful in infrastructure networks like electrical grids that require maintaining connectivity, i.e., stability and robustness against failures [3, 12]. In addition, SC also finds extensive applications in both practical and theoretical fields, including phylogenetics [40], graph sparsification [39], electric circuit analysis [10, 36], and combinatorial optimization [4, 18], to name a few.

Despite its usefulness, the problem of computing the SC values of all edges (AESC) in $\mathcal{G}$ remains challenging. To explain, let $n$ and $m$ be the number of nodes and edges in the graph $\mathcal{G}$, respectively. The graph $\mathcal{G}$ can have $O(n^n)$ spanning trees in the worst case. Hence, the exact AESC computation by enumerating all spanning trees is infeasible. The best-known algorithm [40] for the exact AESC computation is based on Kirchoff's matrix-tree theory [13, 42], which bears more than quadratic time $O(mn^{3/2})$, and thus, is prohibitive for massive graphs. To cope with this challenge, a series of approximation algorithms [14, 29, 34, 39] for AESC have been developed in recent years. Given an absolute error threshold $\epsilon$, existing solutions focus on calculating an estimated SC $\hat{s}(e_{i,j})$ for each edge $e_{i,j}$ with at most $\epsilon$ absolute error in it. Although these methods allow us to trade result accuracy for execution time, they are still rather computationally expensive when $\mathcal{G}$ is sizable and $\epsilon$ is small. Spielman and Srivastava [39] propose to approximate AESC via its equivalent matrix-based definition, leading to $\tilde{O}\left(\frac{m}{\epsilon^2}\right)$ time in total. In the follow-up work [29], Mavroforakis et al. develop a fast implementation by incorporating a suite of heuristic optimizations that considerably elevate its empirical efficiency without compromising its asymptotic performance. However, both methods become impractical when the matrices are high-dimensional and dense (i.e., $n$ and $m$ are large). To sidestep the shortcomings of matrices, Hayashi et al. [14] and Peng et al. [34] capitalize on the idea of using random walks for fast SC estimation, whereas these random walk-based techniques remain $\tilde{O}\left(\frac{m}{\epsilon^2}\right)$ time.

Motivated by the deficiencies of existing solutions, this paper presents two approximation algorithms for AESC: TGT and TGT+. At their hearts lie our improved bounds for random walk truncation, which are obtained through a rigorous theoretical analysis and novel exploitation of eigenvalues/eigenvectors pertaining to $\mathcal{G}$. Notably, compared to Peng et al.'s bound [34], our bound can achieve orders of magnitude of reduction in random walk length. Based thereon, TGT (Truncated Graph Traversal) conducts the graph traversal, i.e., deterministic version of random walks, from each node to probe nodes within the truncated length. In doing so, TGT outperforms the state of the arts in the case where the amount of random walks needed in them exceeds the graph size. To overcome the limitations of TGT on large graphs with high degrees, we further devise TGT+, whose idea is deriving rough estimations of AESC by graph traversals in TGT and refining the results using merely a handful of random walks. By including a greedy trade-off strategy and additional optimizations, we can orchestrate and optimize the entire TGT+ algorithm for enhanced practical efficiency. On the theoretical side, TGT+ propels the approximate AESC computation by improving the asymptotic performance to $\tilde{O}\left(\frac{n}{\epsilon^2} + m\right)$. Our extensive experiments on multiple benchmark graph datasets exhibit that TGT+ is often more than one order of magnitude faster compared to the state-of-the-art solutions while offering uncompromised or even superior result quality. Notably, on the Twitch dataset with 6.8 million edges, TGT+ can achieve $10^{-5}$ empirical error on average within 17 minutes for AESC, using a single CPU core, whereas the best competitor takes over 10 hours.

To summarize, we make the following contributions in this work:
- We derive an improved lower bound for the truncated random walk length and propose a first-cut solution TGT, which estimates AESC using the graph traversal operations. (Section 3)
- We develop an optimized solution TGT+, which integrates random walk sampling into TGT in an adaptive manner and improves over TGT in terms of practical efficiency. (Section 4)
- We compare our proposed solutions with 3 competitors on 5 real datasets and demonstrate the superiority of TGT+. (Section 5)

**Table 1: Frequently used notations.**

| Notation | Description |
| --- | --- |
| $\mathcal{G} = (\mathcal{V}, \mathcal{E})$ | An undirected graph $\mathcal{G}$ with node set $\mathcal{V}$ and edge set $\mathcal{E}$. |
| $n, m$ | The numbers of nodes and edges in $\mathcal{G}$. |
| $\mathcal{N}(v_i), d(v_i)$ | The neighbor set and degree of node $v_i$. |
| $\mathbf{D}, \mathbf{P}$ | The degree and transition matrices of $\mathcal{G}$, respectively. |
| $p_\ell(v_i, v_j)$ | The $\ell$-hop TP, i.e., $\mathbf{P}^\ell[i, j]$. |
| $s(e_{i,j}), \hat{s}(e_{i,j})$ | The exact and estimated SC of edge $e_{i,j}$, respectively. |
| $\tau_{i,j}$ | The truncated length for edge $e_{i,j}$ defined by Eq.(5). |
| $\epsilon, \delta$ | The absolute error threshold and failure probability. |
| $\omega, \gamma$ | The number of eigenvectors and candidate nodes, respectively. |

## 2 PRELIMINARIES

This section sets up the stage for our study by introducing basic notations, the formal problem definition of $\epsilon$-approximate AESC computation, and the main competitors for AESC approximation.

### 2.1 Notations

Let $\mathcal{G} = (\mathcal{V}, \mathcal{E})$ be an undirected graph, where $\mathcal{V}$ is a set of $n$ nodes and $\mathcal{E}$ is a set of $m$ edges. For each edge $e_{i,j} \in \mathcal{E}$, we say $v_i$ and $v_j$ are neighbors to each other, and we use $\mathcal{N}(v_i)$ to denote the set of neighbors of $v_i$, where the degree is $d(v_i) = |\mathcal{N}(v_i)|$. Throughout this paper, we use a boldface lower-case (resp. upper-case) letter $\vec{\mathbf{x}}$ (resp. $\mathbf{M}$) to represent a vector (resp. matrix), with its $i$-th element (resp. element at the $i$-th row and $j$-th column) denoted as $\vec{\mathbf{x}}[i]$ (resp. $\mathbf{M}[i, j]$). Given $\mathcal{G}$, we denote by $\mathbf{A}$ the adjacency matrix of $\mathcal{G}$, where $\mathbf{A}[i, j] = 1$ if $e_{i,j} \in \mathcal{E}$ and $\mathbf{A}[i, j] = 0$ otherwise. In addition, we let $\mathbf{D}$ be the degree diagonal matrix of $\mathcal{G}$ and the diagonal entry $\mathbf{D}[i, i] = d(v_i)$ for each node $v_i \in \mathcal{V}$. Let $\mathbf{P} = \mathbf{D}^{-1}\mathbf{A}$ be the random walk matrix (i.e., transition matrix) of $\mathcal{G}$, in which $\mathbf{P}[i, j] = \frac{1}{d(v_i)}$ if $e_{i,j} \in \mathcal{E}$ and $\mathbf{P}[i, j] = 0$ otherwise. Correspondingly, we denote $p_\ell(v_i, v_j) = \mathbf{P}^\ell[i, j]$, which can be interpreted as the probability of a random walk from node $v_i$ visits node $v_j$ at the $\ell$-th hop, reflecting the proximity of nodes $v_i, v_j$. We refer to $p_\ell(v_i, v_j)$ as $\ell$-hop TP (transition probability) of $v_j$ w.r.t. $v_i$. In this paper, we assume $\mathcal{G}$ is connected and not bipartite. According to [31], the random walks over $\mathcal{G}$ are ergodic, i.e., $\lim_{\ell \to \infty} \mathbf{P}^\ell[i, j] = \frac{d(v_j)}{2m}$ $\forall v_i, v_j \in \mathcal{V}$. Table 1 lists the notations that are frequently used in this paper.

### 2.2 Problem Definition

*Definition 2.1 (Spanning Centrality [40]).* Given an undirected and connected graph $\mathcal{G}$, the SC $s(e_{i,j}) \in (0, 1]$ of an edge $e_{i,j}$ is defined as the fraction of spanning trees of $\mathcal{G}$ that contains $e_{i,j}$.

Definition 2.1 presents the formal definition of SC. Recall that a *spanning tree* of graph $\mathcal{G}$ is a tree and spans over all nodes of $\mathcal{G}$. Intuitively, a high SC $s(e_{i,j})$ quantifies how crucial edge $e_{i,j}$ is for $\mathcal{G}$ to ensure connectedness. Since an edge $e_{i,j}$ with a high SC means that it appears in most spanning trees, all of them will fall apart once $e_{i,j}$ is removed from $\mathcal{G}$. In the extreme case where $s(e_{i,j}) = 1$, $\mathcal{G}$ will be disconnected when $e_{i,j}$ is excluded. To our knowledge, the state-of-the-art algorithm [40] for computing the exact AESC entails $O\left(mn^{\frac{3}{2}}\right)$ time, which is prohibitive for large graphs. Following previous works [14, 34], we focus on $\epsilon$-approximate all-edge SC (AESC) computation, defined as follows. Particularly, we say an estimated SC $\hat{s}(e_{i,j})$ is $\epsilon$-approximate if it satisfies Eq. (1).

*Definition 2.2 ($\epsilon$-Approximate AESC).* Given an undirected and connected graph $\mathcal{G} = (\mathcal{V}, \mathcal{E})$ and an absolute error threshold $\epsilon \in (0, 1)$, the $\epsilon$-approximate AESC computation returns an estimated $\hat{s}(e_{i,j})$ for every edge $e_{i,j} \in \mathcal{E}$ such that

$$|\hat{s}(e_{i,j}) - s(e_{i,j})| \leq \epsilon. \tag{1}$$

## 2.3 State of the Arts

We briefly revisit four recent techniques for AESC computation: Fast-Tree [29], ST-Edge [14], MonteCarlo and MonteCarlo-C [34]. Other related works on SC will be reviewed later in Section 6.

**Fast-Tree.** Mavroforakis et al. [29] develop a fast implementation of [39] on the basis of the equivalence between SC and *effective resistance* (ER) [6] when node pairs are edges. To be more specific, as per [39], the ER of all edges are the diagonal elements of matrix $\mathbf{R} = \mathbf{B}\mathbf{L}^\dagger\mathbf{B}^\top$, where $\mathbf{B}$ and $\mathbf{L}^\dagger$ are the incidence matrix and the pseudoinverse of the Laplacian matrix of $\mathcal{G}$, respectively. Fast-Tree first employs random projections [1] to reduce high matrix dimensions and then deploys the SDD solver to solve the linear systems in the low-dimensional space, resulting in a linear time complexity of $O\left(\frac{m}{\epsilon^2} \log^2 n \log\left(\frac{1}{\epsilon}\right)\right)$. However, its practical efficiency is less than satisfactory on large graphs, as revealed by the experiments in [14].

**ST-Edge.** Based on Definition 2.1, Hayashi et al. [14] first sample a sufficient number of random spanning trees by Wilson's algorithm [48], and record the fraction of trees where edge $e_{i,j}$ appears as the estimated $\hat{s}(e_{i,j})$. As proved, the expected time to draw a spanning tree rooted at a random node $v_r$ is $O\left(\sum_{v_i \in \mathcal{V}} \frac{d(v_i)}{m} \kappa(v_i, v_r)\right)$, where $\kappa(v_i, v_r)$ is the commute time between nodes $v_i$ and $v_r$ and is $O(m)$ [31]. Hence, to ensure the $\epsilon$-approximate for estimated AESC values, ST-Edge runs in $O\left(\frac{m}{\epsilon^2} \log(\frac{m}{\delta})\right)$ time by sampling $O\left(\frac{1}{\epsilon^2} \log(\frac{m}{\delta})\right)$ spanning trees, rendering it costly when $\epsilon$ is small.

**MonteCarlo and MonteCarlo-C.** Very recently, Peng et al. [34] theoretically establish another equivalent definition of SC $s(e_{i,j})$:

$$s(e_{i,j}) = \sum_{\ell=0}^{\infty} \frac{p_\ell(v_i, v_i)}{d(v_i)} + \frac{p_\ell(v_j, v_j)}{d(v_j)} - \frac{p_\ell(v_i, v_j)}{d(v_j)} - \frac{p_\ell(v_j, v_i)}{d(v_i)}.$$

Thus, the problem is transformed into computing $\ell$-hop TP values of every two nodes in $\{v_i, v_j\}$ for $0 \leq \ell \leq \infty$. The crux of MonteCarlo and MonteCarlo-C involves finding a truncated length $\tau$ for random walks, which ensures $|s_\tau(e_{i,j}) - s(e_{i,j})| \leq \frac{\epsilon}{2}$, where

$$s_\tau(e_{i,j}) = \sum_{\ell=0}^{\tau} \frac{p_\ell(v_i, v_i)}{d(v_i)} + \frac{p_\ell(v_j, v_j)}{d(v_j)} - \frac{p_\ell(v_i, v_j)}{d(v_j)} - \frac{p_\ell(v_j, v_i)}{d(v_i)}. \tag{2}$$

Based thereon, MonteCarlo and MonteCarlo-C simulate random walks with length at most $\tau$ from $v_i, v_j$ to approximate the $\ell$-hop TP values such that $|\hat{s}(e_{i,j}) - s_\tau(e_{i,j})| \leq \frac{\epsilon}{2}$ holds, connoting that $\hat{s}(e_{i,j})$ is $\epsilon$-approximate. In particular, Peng et al. [34] provides the following bound for $\tau$ to ensure the $\epsilon$-approximation

$$\tau \geq \left\lceil \log\left(\frac{4}{\epsilon - \epsilon\lambda}\right) / \log\left(\frac{1}{\lambda}\right) - 1 \right\rceil, \tag{3}$$

where $\lambda$ is matrix $\mathbf{P}$'s second largest eigenvalue in absolute value.

The major distinction between MonteCarlo and MonteCarlo-C lies in the approach to computing $\ell$-hop TP values. Specifically, MonteCarlo simply conducts random walks of length $\ell$ ($1 \leq \ell \leq \tau$) to approximate $\ell$-hop TP values before aggregating them as the estimated SC. According to the Chernoff-Hoeffding bound, a total time complexity of $O\left(\frac{\tau^3 \log(8\tau/\delta)}{\epsilon^2}\right)$ is needed to obtain an $\epsilon$-approximate SC $\hat{s}(e_{i,j})$ with a success probability at least $1 - \delta$. By contrast, MonteCarlo-C regards the $\ell$-hop TP $p_\ell(v_i, v_j)$ ($1 \leq \ell \leq \tau$) as the collision probability of two random walks of length-$\frac{\ell}{2}$ from $v_i$ and $v_j$, respectively, and then samples $40000 \times \left(\tau\sqrt{\tau\beta_\ell}/\epsilon + \tau^3 \beta_\ell^{3/2}/\epsilon^2\right)$ length-$(\ell/2)$ random walks from respective nodes. The parameter $\beta_\ell$ is a constant depending on the graph structure, which is hard to compute in practice. Notice that both algorithms are originally designed for computing the ER of any node pair in $\mathcal{G}$, which overlook the unique property of edges and thus are not optimized for AESC computation. Moreover, they require an exorbitant amount of random walks due to the large $\tau$ (up to thousands when $\epsilon$ is small), significantly exacerbating the efficiency issues.

## 3 THE TGT ALGORITHM

In this section, we propose TGT, an iterative deterministic graph traversal approach to AESC processing based on the idea of computing the truncated SC (Eq. (2)) as in MonteCarlo. Particularly, TGT improves over MonteCarlo in two aspects. First and foremost, TGT offers significantly superior edge-wise lower bounds for truncated lengths by leveraging the well-celebrated theory of Markov chains [47] (Section 3.1). Further, TGT develops a deterministic graph traversal method to remedy the efficiency issue caused by substantial random walks needed in MonteCarlo (Section 3.2).

### 3.1 Improved Bounds for Truncated Lengths

LEMMA 3.1 ([47]). *Given an undirected graph $\mathcal{G}$, let $1 = |\lambda_1| \geq |\lambda_2| \geq \cdots \geq |\lambda_n| \geq 0$ be the sorted absolute eigenvalues of $\mathbf{D}^{\frac{1}{2}}\mathbf{P}\mathbf{D}^{-\frac{1}{2}}$ and $\boldsymbol{\varphi}_1, \boldsymbol{\varphi}_2, \ldots, \boldsymbol{\varphi}_n$ be their corresponding normalized eigenvectors. Then, for any two nodes $v_i, v_j \in \mathcal{V}$ and any integer $\ell \geq 0$, we have*

$$\frac{p_\ell(v_i, v_j)}{d(v_j)} = \frac{p_\ell(v_j, v_i)}{d(v_i)} = \frac{1}{2m} \sum_{k=1}^{n} \vec{\mathbf{f}}_k[i] \cdot \vec{\mathbf{f}}_k[j] \cdot \lambda_k^\ell, \tag{4}$$

*where $\vec{\mathbf{f}}_i = \sqrt{2m} \cdot \mathbf{D}^{-\frac{1}{2}} \boldsymbol{\varphi}_i$ for $i = 1, 2, \ldots, n$, and $\vec{\mathbf{f}}_1$ is taken to be $\mathbf{1}$.*

By Lemma 3.1, the $\ell$-hop TP $p_\ell(v_i, v_j)$ can be computed based on the eigenvectors and eigenvalues of matrix $\mathbf{D}^{\frac{1}{2}}\mathbf{P}\mathbf{D}^{-\frac{1}{2}}$, and hence, the difference between $s_\tau(e_{i,j})$ and $s(e_{i,j})$ can be quantified via

$$|s_\tau(e_{i,j}) - s(e_{i,j})| = \left| \sum_{\ell=\tau+1}^{\infty} \frac{1}{2m} \sum_{k=1}^{n} (\vec{\mathbf{f}}_k[i] - \vec{\mathbf{f}}_k[j])^2 \lambda_k^\ell \right|.$$

This suggests that we can utilize these eigenvectors and eigenvalues to determine a truncated length $\tau_{i,j}$ for edge $e_{i,j}$ so that $|s_\tau(e_{i,j}) - s(e_{i,j})| \leq \frac{\epsilon}{2}$. Additionally, when $\ell = 1$ and $e_{i,j} \in \mathcal{E}$, we have $\frac{2m}{d(v_i) \cdot d(v_j)} = \sum_{k=1}^{n} \vec{\mathbf{f}}_k[i] \cdot \vec{\mathbf{f}}_k[j] \cdot \lambda_k$ as per Eq. (4). Given the above observations, we can establish an improved lower bound for the truncated length $\tau_{i,j}$ of each edge $e_{i,j}$, as shown in Theorem 3.2. For ease of exposition, we defer all proofs to Appendix A.

THEOREM 3.2. *Given $\mathcal{G} = (\mathcal{V}, \mathcal{E})$, $|s(e_{i,j}) - s_\tau(e_{i,j})| \leq \epsilon$ holds for any edge $e_{i,j} \in \mathcal{E}$ when $\tau_{i,j}$ satisfies*

$$\tau_{i,j} \geq f(e_{i,j}, \epsilon) \text{ and } \tau_{i,j} \equiv 1 \pmod{2} \tag{5}$$

**Algorithm 1:** CalTau

**Data:** Graph $\mathcal{G}$, $\{\lambda_1, \ldots, \lambda_\omega\}$, $\{\vec{f}_1, \ldots, \vec{f}_\omega\}$
**Parameters:** $e_{i,j}$, $\epsilon$
**Result:** $\tau_{i,j}$

1 $\tau_{i,j} \leftarrow$ Eq. (6) with $\lambda_2$ and $\Delta_t = \Upsilon = 0$;
2 $\Upsilon \leftarrow$ Eq. (7); $t \leftarrow 1$;
3 **while** true **do**
4 $\quad \Delta_t \leftarrow$ Eq. (8); $\tau' \leftarrow$ Eq. (6);
5 $\quad$ **if** $t \leq \tau'$ **then** $\tau_{i,j} \leftarrow \tau'$; $t \leftarrow t + 2$ ;
6 $\quad$ **else** break;
7 **return** $\tau_{i,j} \leftarrow t$;

$$f(e_{i,j}, \epsilon) = \left\lceil \frac{\log\left(\frac{\frac{1}{d(v_i)} + \frac{1}{d(v_j)} - \frac{2}{d(v_i) \cdot d(v_j)} - \Upsilon}{(\epsilon - \Delta_t) \cdot (1 - \lambda_\omega^2)}\right)}{\log\left(\frac{1}{|\lambda_\omega|}\right)} - 1 \right\rceil, \quad (6)$$

where

$$\Upsilon = \frac{1}{2m} \sum_{k=2}^{\omega-1} (\vec{f}_k[i] - \vec{f}_k[j])^2 \cdot (1 + \lambda_k), \quad (7)$$

$$\Delta_t = \frac{1}{2m} \sum_{k=2}^{\omega-1} (\vec{f}_k[i] - \vec{f}_k[j])^2 \cdot \frac{\lambda_k^{t+1}}{1 - \lambda_k}, \quad (8)$$

and $t$ is an odd number ensuring $t \leq \tau_{i,j}$.

Compared to Peng et al.'s $\tau$ in Eq. (3), our truncated length $\tau_{i,j}$ of edge $e_{i,j}$ in Theorem 3.2 is dependent to the degrees of nodes $v_i, v_j$, the $\omega$-largest (typically $\omega = 128$) eigenvalues in absolute value and their corresponding eigenvectors of $\mathbf{D}^{\frac{1}{2}} \mathbf{P} \mathbf{D}^{-\frac{1}{2}}$, enabling up to orders of magnitude improvement in practice, as reported in Figure 1. Note that the eigenvalues and eigenvectors can be efficiently computed in the preprocessing stage (see Figure 4).

Algorithm 1 presents the pseudo-code of CalTau, an algorithm realizing the computation of $\tau_{i,j}$ on the basis of Theorem 3.2. Given graph $\mathcal{G}$, $\omega$ eigenvalues $\{\lambda_1, \ldots, \lambda_\omega\}$, eigenvectors $\{\vec{f}_1, \ldots, \vec{f}_\omega\}$, and parameters $e_{i,j}$, $\epsilon$ as inputs, CalTau initializes $\tau_{i,j}$ by Eq. (6) with $\lambda_2$ and $\Upsilon = \Delta_t = 0$ at Line 1, followed by setting $t = 1$ and calculating $\Upsilon$ according to Eq. (7) at Line 2. After that, CalTau increases $t$ iteratively to search for the optimal $t$ such that it is closest to but does not exceed Eq. (6), ensuring the validity of Theorem 3.2 (Lines 3–6). To be more precise, in each iteration, CalTau calculates a candidate truncated length $\tau'$ using Eq. (6), wherein $\Delta_t$ is obtained by Eq. (8) with current $t$. Next, if $t \leq \tau'$, we update $\tau_{i,j}$ as $\tau'$ and increase $t$ by 2 (Line 5). CalTau repeats the above procedure until the condition at Line 5 does not hold and returns $t$ as $\tau_{i,j}$ at Line 7.

### 3.2 Complete Algorithm and Analysis

In light of Theorem 3.2, the problem of AESC computation in Definition 2.2 is reduced to computing the approximate SC $\hat{s}(e_{i,j}) = s_\tau(e_{i,j})$ as per Eq. (2) for each edge $e_{i,j} \in \mathcal{E}$. Unlike prior methods, TGT conducts a deterministic graph traversal from each node $v_i \in \mathcal{V}$ to compute $\ell$-hop TP values $p_\ell(v_i, v_i)$ and $p_\ell(v_j, v_i)$ for $1 \leq \ell \leq \tau_{i,j}$

**Algorithm 2:** TGT

**Data:** Graph $\mathcal{G}$
**Parameters:** $\epsilon$
**Result:** $s_\tau(e_{i,j}) \ \forall e_{i,j} \in \mathcal{E}$

1 **for** $v_i \in \mathcal{V}$ **do**
2 $\quad p_0(v_j, v_i) \leftarrow 0 \ \forall v_j \in \mathcal{V} \setminus v_i$; $p_0(v_i, v_i) \leftarrow 1$;
3 $\quad g_\tau(v_i, v_j) \leftarrow \frac{1}{d(v_i)} \ \forall v_j \in \mathcal{N}(v_i)$;
4 $\quad \tau_p \leftarrow \max_{v_j \in \mathcal{N}(v_i)}$ CalTau$(e_{i,j}, \epsilon)$;
5 $\quad$ **for** $\ell \leftarrow 1$ to $\tau_p$ **do**
6 $\quad\quad p_\ell(v_j, v_i) \leftarrow 0 \ \forall v_j \in \mathcal{V}$;
7 $\quad\quad$ **for** $v_j \in \mathcal{V}$ with $p_{\ell-1}(v_j, v_i) > 0$ **do**
8 $\quad\quad\quad$ **for** $v_x \in \mathcal{N}(v_j)$ **do**
9 $\quad\quad\quad\quad p_\ell(v_x, v_i) \leftarrow p_\ell(v_x, v_i) + \frac{p_{\ell-1}(v_j, v_i)}{d(v_x)}$;
10 $\quad\quad$ **for** $v_j \in \mathcal{N}(v_i)$ **do**
11 $\quad\quad\quad g_\tau(v_i, v_j) \leftarrow g_\tau(v_i, v_j) + \frac{p_\ell(v_i, v_i)}{d(v_i)} - \frac{p_\ell(v_j, v_i)}{d(v_i)}$;
12 **for** $e_{i,j} \in \mathcal{E}$ **do** $s_\tau(e_{i,j}) \leftarrow g_\tau(v_i, v_j) + g_\tau(v_j, v_i)$;
13 **return** $s_\tau(e_{i,j}) \ \forall e_{i,j} \in \mathcal{E}$;

in an iterative manner, and aggregates them as

$$g_\tau(v_i, v_j) = \sum_{\ell=0}^{\tau} \frac{p_\ell(v_i, v_i)}{d(v_i)} - \sum_{\ell=0}^{\tau} \frac{p_\ell(v_j, v_i)}{d(v_i)} \quad (9)$$

to further derive $s_\tau(e_{i,j})$ by $g_\tau(v_i, v_j) + g_\tau(v_j, v_i)$. The pseudo-code of TGT is illustrated in Algorithm 2. In the course of graph traversal from each node $v_i \in \mathcal{V}$ (Lines 2–11), $p_0(v_i, v_j)$ and $g_\tau(v_i, v_j) \ \forall v_j \in \mathcal{V}$ are initialized as Lines 2–3. Afterward, at Line 4, TGT invokes Algorithm 1 with absolute error $\epsilon$ and each edge $e_{i,j}$ that is adjacent to $v_i$. Let $\tau_p$ be the largest truncated length $\tau_{i,j}$ for all $v_j \in \mathcal{N}(v_i)$. Then, Algorithm 2 performs a $\tau_p$-hop graph traversal originating from $v_i$ (Lines 5–11). Specifically, at $\ell$-th hop, TGT first sets $p_\ell(v_j, v_i) = 0 \ \forall v_j \in \mathcal{V}$. Subsequently, for each node $v_j$ with non-zero $(\ell-1)$-hop TP $p_{\ell-1}(v_j, v_i)$, we scatter its value to its neighbors, i.e., visit each neighbor $v_x \in \mathcal{N}(v_j)$ by adding $\frac{p_{\ell-1}(v_j, v_i)}{d(v_x)}$ to $p_\ell(v_x, v_i)$. This operation essentially performs a sparse matrix-vector multiplication $p_\ell(\cdot, v_i) = \mathbf{P} \cdot p_{\ell-1}(\cdot, v_i)$. With $p_\ell(v_x, v_i) \ \forall v_x \in \mathcal{V}$, TGT injects an increment of $\frac{p_\ell(v_i, v_i)}{d(v_i)} - \frac{p_\ell(v_j, v_i)}{d(v_i)}$ to $g_\tau(v_i, v_j)$ for each neighbor $v_j$ of $v_i$. After the completion of all graph traversal operations, Algorithm 2 computes $s_\tau(e_{i,j})$ for each edge $e_{i,j}$ (Line 12) and returns them as the answers. The following theorem states the correctness and the worst-case time complexity of TGT.

THEOREM 3.3. *Algorithm 2 returns $\epsilon$-approximate* SC *values $s_\tau(e_{i,j})$ $\forall e_{i,j} \in \mathcal{E}$ using $O\left(nm \log\left(\frac{1}{\epsilon}\right)\right)$ time in the worst case.*

Notwithstanding its unsatisfying *worst-case* time complexity, by virtue of our improved lower bounds for truncated lengths in Section 3.1, the actual number of graph traversal operations from each node in Algorithm 2 (Lines 7–9) is far less than $O(m)$ when $\epsilon$ is non-diminutive, strengthening the superiority of TGT over MonteCarlo in empirical efficiency.

# 4 THE TGT+ ALGORITHM

Although TGT advances MonteCarlo in practical performance, we observe in our experiments that its cost is intolerable for massive graphs with high degrees. The reason is that the number of non-zero $\ell$-hop TP values grows explosively at an astonishing rate till $m$ (Lines 7–9 in Algorithm 2) on such graphs as $\ell$ increases, causing a quadratic computational complexity of $O(nm)$. The severity of the efficiency issue is accentuated in high-precision AESC computation, i.e., $\epsilon$ is small. To alleviate this issue, we propose TGT+, an algorithm that significantly improves TGT in terms of both practical efficiency and asymptotic performance. The rest of this section proceeds as follows: Section 4.1 delineates the basic idea of TGT+, followed by several optimization techniques in Section 4.2. Finally, Section 4.3 describes the complete algorithm and analysis.

## 4.1 High-level Idea

Considering the sheer number of non-zero $\ell$-hop TP values in TGT when $\ell$ is increased, we propose to calculate the TP values within $\tilde{\tau}$ (a small number) hops using TGT and harness random walks for the estimation of $\ell$-hop TP with $\ell > \tilde{\tau}$. The rationale is that the amount of nodes in the vicinity of a given node $v_i$ is often limited, and hence, can be efficiently covered by a graph traversal from $v_i$. On the contrary, far-reaching nodes from $v_i$ can be multitudinous (up to millions in large graphs), where random walks suit the demand better by focusing on probing *important* nodes (i.e., with high TP values) in lieu of all of them. To fulfill the above-said idea, we first derive a truncated length $\tau_{i,j}$ such that $|s(e_{i,j}) - s_\tau(e_{i,j})| \le \frac{\epsilon}{2}$ for each edge $e_{i,j} \in \mathcal{E}$. Next, the problem is computing an estimated SC $\hat{s}(e_{i,j})$ of each edge $e_{i,j}$ to ensure $|\hat{s}(e_{i,j}) - s_\tau(e_{i,j})| \le \frac{\epsilon}{2}$ using graph traversals and random walks. To facilitate the seamless integration of random walks into TGT, we leverage the following crucial property of $g_\tau(v_i, v_j)$, a constituent part of SC $s_\tau(e_{i,j})$ as defined in Eq. (9).

**LEMMA 4.1.** *For any integer $\tau$ and $\tilde{\tau}$ with $0 \le \tilde{\tau} \le \tau$,*

$$g_\tau(v_i, v_j) = g_{\tilde{\tau}}(v_i, v_j) + g_{\tilde{\tau} \to \tau}(v_i, v_j),$$

*where*

$$g_{\tilde{\tau} \to \tau}(v_i, v_j) = \sum_{v_x \in V} \frac{p_{\tilde{\tau}}(v_x, v_i)}{d(v_i)} \left( \sum_{\ell=1}^{\tau_{i,j} - \tilde{\tau}} p_\ell(v_i, v_x) - p_\ell(v_j, v_x) \right). \quad (10)$$

More concretely, given a cherry-picked length $\tilde{\tau}$ ($1 \le \tilde{\tau} \le \tau_{i,j}$), Lemma 4.1 implies that we can estimate $g_{\tilde{\tau} \to \tau}(v_i, v_j)$ by simulating random walks of lengths from 1 to $\tau_{i,j} - \tilde{\tau}$ after obtaining $g_{\tilde{\tau}}(v_i, v_j)$ and $p_{\tilde{\tau}}(\cdot, v_i)$ with TGT. Mathematically, if we conduct two length-$(\tau_{i,j} - \tilde{\tau})$ random walks $W_i$ and $W_j$ containing visited nodes from nodes $v_i, v_j$, respectively, we can define a random variable $X$ as

$$X = \frac{1}{d(v_i)} \cdot \left( \sum_{v_x \in W_i} p_{\tilde{\tau}}(v_x, v_i) - \sum_{v_y \in W_j} p_{\tilde{\tau}}(v_y, v_i) \right). \quad (11)$$

By definition, the expectation $\mathbb{E}[X]$ of $X$ is exactly $g_{\tilde{\tau} \to \tau}(v_i, v_j)$ in Eq. (10), indicating that $X$ is an unbiased estimator of $g_{\tilde{\tau} \to \tau}(v_i, v_j)$. Suppose that the range of $X$ is bounded by

$$|X| \le \frac{\chi}{d(v_i)}. \quad (12)$$

---

**Algorithm 3:** CalChi

**Data:** Graph $\mathcal{G}$, edge $(v_i, v_j)$, $p_{\tilde{\tau}}(\cdot, v_i)$
**Parameters:** $\gamma$
**Result:** $\chi$

1  **if** $\gamma > 0$ **then**
2  　Identify $C = \{c_1, c_2, \ldots, c_\gamma\}$ such that $p_{\tilde{\tau}}(c_1, v_i) \ge p_{\tilde{\tau}}(c_2, v_i) \ge \cdots \ge p_{\tilde{\tau}}(c_\gamma, v_i) \ge p_{\tilde{\tau}}(v_l, v_i) \; \forall v_l \in \mathcal{V} \setminus C$;
3  　**if** $\exists c_a, c_b \in C$ and $(c_a, c_b) \in \mathcal{E}$ **then**
4  　　$\hat{\rho}_i \leftarrow \max_{(c_a, c_b) \in \mathcal{E}, \forall c_a, c_b \in C} p_{\tilde{\tau}}(c_a, v_i) + p_{\tilde{\tau}}(c_b, v_i)$;
5  　**else** $\hat{\rho}_i \leftarrow p_{\tilde{\tau}}(c_1, v_i) + p_{\tilde{\tau}}(c_\gamma, v_i)$;
6  　$\chi \leftarrow$ Eq. (17) with $\rho_i = \hat{\rho}_i$;
7  **else** $\chi \leftarrow$ Eq. (15);
8  **return** $\chi$;

---

**LEMMA 4.2 (HOEFFDING'S INEQUALITY [15]).** *Let $Z_1, Z_2, \ldots, Z_{n_z}$ be independent random variables with $Z_i$ ($\forall 1 \le i \le n_z$) is strictly bounded by the interval $[a_j, b_j]$. We define the empirical mean of these variables by $Z = \frac{1}{n_z} \sum_{i=1}^{n_z} Z_i$. Then,*

$$\mathbb{P}[|Z - \mathbb{E}[Z]| \ge \epsilon] \le 2 \exp\left(-\frac{2n_z^2 \epsilon^2}{\sum_{j=1}^{n_z}(b_j - a_j)^2}\right).$$

It is straightforward to apply Hoeffding's inequality in Lemma 4.2 to derive the total number of random walks needed for the accurate estimation of $g_{\tilde{\tau} \to \tau}(v_i, v_j)$, i.e.,

$$R(e_{i,j}, \tau_{i,j} - \tilde{\tau}) = \frac{8\chi^2 \log\left(\frac{2m}{\delta}\right)}{d(v_i)^2 \cdot \epsilon^2}. \quad (13)$$

In the subsequent section, we elucidate the determination of $\tilde{\tau}$ and $\chi$ so as to strike a good balance between graph traversal and random walks for optimized performance and meanwhile reduce the number $R(e_{i,j}, \tau_{i,j} - \tilde{\tau})$ of samples required.

## 4.2 Optimizations

*4.2.1 Adaptive determination of $\tilde{\tau}$.* Since the length of random walks for estimating $g_{\tilde{\tau} \to \tau}(v_i, v_j) \; \forall v_j \in \mathcal{N}(v_i)$ is $\tau_{i,j} - \tilde{\tau}$, the computational overhead incurred by random walks from the given node $v_i$ and its neighbors is hence bounded by

$$O\left(\sum_{v_j \in \mathcal{N}(v_i)} R(e_{i,j}, \tau_{i,j} - \tilde{\tau}) \cdot (\tau_{i,j} - \tilde{\tau})\right),$$

which increases as $\tilde{\tau}$ decreases. Conversely, the graph traversal operations in TGT will reduce considerably when $\tilde{\tau}$ is lowered, as explained at the beginning of Section 4.1. In short, the length $\tilde{\tau}$ controls the trade-off between the deterministic graph traversal and random walks for each node $v_i \in \mathcal{V}$. Since it is hard to accurately quantify the graph traversal cost as a function regarding $\tilde{\tau}$ due to the complex graph structure, we make use of an adaptive strategy to determine $\tilde{\tau}$. More precisely, in the $\ell$-th iteration of deterministic graph traversal (Lines 6-9 in Algorithm 2) originating from $v_i$, we set $\tilde{\tau} = \ell$ and switch the graph traversal to random walk simulations, if the following inequality holds:

$$\sum_{v_j \in \mathcal{V} \& p_\ell(v_j, v_i) \neq 0} d(v_j) > \sum_{v_j \in \mathcal{N}(v_i)} R(e_{i,j}, \tau_{i,j} - \ell), \quad (14)$$

where the l.h.s. and r.h.s. represent their respective costs for computing $(\ell + 1)$-hop TP values in the next iteration. The rationale of Eq. (14) is that we choose random walks rather than the graph traversal when the cost of the latter will outstrip the former.

### 4.2.2 Effective refinement of $\chi$.
By the definition of $X$ in Eq. (11), one may simply set $\chi$ as follows:

$$\chi = 2 \cdot (\tau - \tilde{\tau}) \cdot \left( \max_{v_x \in V} p_{\tilde{\tau}}(v_x, v_i) - \min_{v_y \in V} p_{\tilde{\tau}}(v_y, v_i) \right), \quad (15)$$

where $p_{\tilde{\tau}}(v_x, v_i) \; \forall v_x \in \mathcal{V}$ is known from TGT. Unfortunately, the empirical values of r.h.s. of Eq. (15) are usually ineligible on real graphs, resulting in a considerable number of random samples according to Eq. (13). Intuitively, given that $W_i$ and $W_j$ are the random walks from two adjacent nodes $v_i, v_j$ (i.e., $e_{i,j} \in \mathcal{E}$), respectively, the nodes on $W_i$ and $W_j$ are highly overlapped. As a consequence, the difference between $\sum_{v_x \in W_i} p_{\tilde{\tau}}(v_x, v_i)$ and $\sum_{v_y \in W_j} p_{\tilde{\tau}}(v_y, v_i)$ in Eq. (11) (i.e., $\chi$) would be insignificant in practice. Inspired by the aforementioned observation, we can establish the following lower and upper bounds for $\sum_{v_x \in W_i} p_{\tilde{\tau}}(v_x, v_i)$.

LEMMA 4.3. *Let $W_i$ be any length-$\ell$ ($\ell \geq 1$) random walk over $\mathcal{G}$ starting from node $v_i$ and $\rho_i$ be*

$$\rho_i = \max_{e_{x,y} \in \mathcal{E}} \{ p_{\tilde{\tau}}(v_x, v_i) + p_{\tilde{\tau}}(v_y, v_i) \}. \quad (16)$$

*Then, we have $LB(v_i, \ell) \leq \sum_{v_x \in W_i} p_{\tilde{\tau}}(v_x, v_i) \leq UB(v_i, \ell)$, where the lower and upper bounds $LB(v_i, \ell), UB(v_i, \ell)$ are defined by*

$$LB(v_i, \ell) = \min_{v_l \in \mathcal{N}(v_i)} \{ p_{\tilde{\tau}}(v_l, v_i) \} + (\ell - 1) \cdot \min_{v_l \in \mathcal{V}} \{ p_{\tilde{\tau}}(v_l, v_i) \}, \text{ and}$$

$$UB(v_i, \ell) = \max_{v_l \in \mathcal{N}(v_i)} \left\{ \frac{p_{\tilde{\tau}}(v_l, v_i)}{2} \right\} + \max_{v_l \in \mathcal{V}} \left\{ \frac{p_{\tilde{\tau}}(v_l, v_i)}{2} \right\} + \frac{(\ell - 1) \cdot \rho_i}{2}.$$

Using Lemma 4.3, a refined $\chi$ is at hand:

$$\chi = UB(v_i, \tau_{i,j} - \tilde{\tau}) + UB(v_j, \tau_{i,j} - \tilde{\tau}) \\ - LB(v_i, \tau_{i,j} - \tilde{\tau}) - LB(v_j, \tau_{i,j} - \tilde{\tau}). \quad (17)$$

It is worth mentioning that $LB(v_i, \ell)$ and the first two terms in $UB(v_i, \ell)$ can be efficiently computed without sorting all the $n$ nodes, since the actual number of non-zero entries in $p_{\tilde{\tau}}(\cdot, v_i)$ is limited due to our fine-tuned $\tilde{\tau}$, as remarked earlier. Therefore, the critical challenge to realize the derivation of the improved $\chi$ in Eq. (17) arises from the computation of $\rho_i$ in Eq. (16), which incurs a high cost of $O(m \log m)$ if we search for the optimal edge $e_{x,y}$ ensuring Eq. (16) from $\mathcal{E}$ in a brute-force fashion. To tackle this problem, we propose a subroutine CalChi in Algorithm 3, which computes $\rho_i$ for $\chi$ in a cost-effective manner, without jeopardizing its correctness. More specifically, instead of inspecting all the $m$ edges in $\mathcal{G}$, CalChi first identifies a set $C = \{c_1, c_2, \cdots, c_\gamma\}$ of nodes from $\mathcal{V}$ with $\gamma$ ($\gamma$ is a small constant) largest $\tilde{\tau}$-hop TP values to $v_i$, in other words $p_{\tilde{\tau}}(c_1, v_i) \geq p_{\tilde{\tau}}(c_2, v_i) \geq \cdots \geq p_{\tilde{\tau}}(c_\gamma, v_i) \geq p_{\tilde{\tau}}(v_l, v_i) \; \forall v_l \in \mathcal{V} \setminus C$ (Line 2). After that, CalChi checks if any two nodes in $C$ form an edge. If $C$ does not contain such two nodes $(c_a, c_b) \in \mathcal{E}$, we set $\rho_i$'s upper bound $\hat{\rho}_i$ to $p_{\tilde{\tau}}(c_1, v_i) + p_{\tilde{\tau}}(c_\gamma, v_i)$, otherwise we use the largest $p_{\tilde{\tau}}(c_a, v_i) + p_{\tilde{\tau}}(c_b, v_i)$ among all edges of $C \times C$, which

---

**Algorithm 4:** TGT+

**Data:** Graph $\mathcal{G}$
**Parameters:** $\epsilon, \delta, \gamma$
**Result:** $\hat{s}(e_{i,j}) \; \forall e_{i,j} \in \mathcal{E}$

1 **for** $v_i \in V$ **do**
2     $\forall v_j \in \mathcal{N}(v_i) \; \tau_{i,j} \leftarrow \text{CalTau}(e_{i,j}, \frac{\epsilon}{2}); \; \ell \leftarrow 0;$
    Lines 3-4 are the same as Lines 2-3 in Algorithm 2;
5     **while** *Eq. (14) fails* **do**
       Lines 6-9 are the same as Lines 6-9 in Algorithm 2;
10        $\ell \leftarrow \ell + 1;$
11     $\tilde{\tau} \leftarrow \ell;$
12     **for** $v_j \in \mathcal{N}(v_i)$ *such that* $\tau_{i,j} - \tilde{\tau} > 0$ **do**
13        $\chi \leftarrow \text{CalChi}(\mathcal{G}, v_i, v_j, p_{\tilde{\tau}}(\cdot, v_i), \gamma);$
14        $n_r \leftarrow R(e_{i,j}, \tau_{i,j} - \tilde{\tau})$ in Eq. (13);
15        **for** $i \leftarrow 1$ *to* $n_r$ **do**
16           Simulate two length-$(\tau_{i,j} - \tilde{\tau})$ random walks $W_i$, $W_j$ from $v_i, v_j$, respectively;
17           $\hat{g}_\tau(v_i, v_j) \leftarrow \hat{g}_\tau(v_i, v_j) + \frac{X}{n_r}$ with $X$ = Eq. (11);
18 **for** $e_{i,j} \in \mathcal{E}$ **do** $\hat{s}(e_{i,j}) \leftarrow \hat{g}_\tau(v_i, v_j) + \hat{g}_\tau(v_j, v_i);$
19 **return** $\hat{s}(e_{i,j}) \; \forall e_{i,j} \in \mathcal{E};$

---

is $\rho_i$ itself (Lines 3-5). The rationale is that when no edges exists in $C \times C$, at least an endpoint $v_y$ of the desired edge $e_{x,y}$ is outside $C$, meaning $p_{\tilde{\tau}}(v_y, v_i) \leq p_{\tilde{\tau}}(c_\gamma, v_i)$. In the meantime, another endpoint $v_x$ of $e_{x,y}$ satisfies $p_{\tilde{\tau}}(v_x, v_i) \leq p_{\tilde{\tau}}(c_1, v_i)$. Accordingly, $p_{\tilde{\tau}}(c_1, v_i) + p_{\tilde{\tau}}(c_\gamma, v_i)$ can serve as an upper bound of $\rho_i$ in this case. Eventually, CalChi calculates $\chi$ according to Eq. (17) by replacing $\rho_i$ by its upper bound $\hat{\rho}_i$ (Line 6). Particularly, when $\gamma = 0$, CalChi degrades to computing $\chi$ by Eq. (15).

## 4.3 Complete Algorithm and Analysis

Algorithm 4 summarizes the procedure of TGT+, which begins with computing $g_\tau(v_i, v_j)$ for each node $v_i \in \mathcal{V}$ and each of its neighbors $v_j$ as in TGT. Specifically, for each node $v_i \in \mathcal{V}$, TGT+ first computes $\tau_{i,j}$ for each neighbor $v_j$ of $v_i$ by taking $\epsilon/2$ as input (Line 2). Subsequently, TGT+ carries out graph traversals as illustrated in TGT for the computation of $g_{\tilde{\tau}}(v_i, v_j) \; \forall v_j \in \mathcal{N}(v_i)$ and $p_{\tilde{\tau}}(\cdot, v_i)$ (Lines 3-10). The iterative process of the graph traversal terminates when Eq. (14) holds and Algorithm 4 then proceeds to sampling random walks for each neighboring node $v_j$ of $v_i$ whose $g_{\tilde{\tau}}(v_i, v_j)$ is insufficiently accurate, i.e., $\tau_{i,j} > \tilde{\tau}$ (Line 12). In particular, TGT+ first invokes Algorithm 3 to obtain the refined $\chi$ (Line 13) before determining the number of random walks $n_r$ at Line 14. Afterwards, TGT+ generates $n_r$ length-$(\tau_{i,j} - \tilde{\tau})$ random walks $W_i, W_j$ from nodes $v_i, v_j$, respectively (Lines 15-16). After each sampling, it increases $\hat{g}_\tau(v_i, v_j)$ by $\frac{X}{n_r}$, where $X$ is a random variable based on Eq. (11) (Line 17). In the end, TGT+ computes $\hat{s}(e_{i,j}) = \hat{g}_\tau(v_i, v_j) + \hat{g}_\tau(v_j, v_i)$ for each edge $e_{i,j} \in \mathcal{E}$ and outputs them as the SC estimations. The following theorem expresses the correctness and complexity of it.

THEOREM 4.4. *For any $\epsilon, \delta \in (0, 1)$, Algorithm 4 returns the $\epsilon$-approximate SC $\hat{s}(e_{i,j}) \; \forall e_{i,j} \in \mathcal{E}$ with the probability at least $1 - \delta$, using $O\left( \frac{1}{\epsilon^2} \cdot \log^3(\frac{1}{\epsilon}) \cdot \log(\frac{m}{\delta}) \cdot \sum_{v_i \in \mathcal{V}} \frac{1}{d(v_i)} + m \right)$ expected time.*

Table 2: Algorithms for $\epsilon$-approximate AESC computation.

| Algorithm | Time Complexity |
|---|---|
| Fast-Tree [29] | $O\left(\frac{m}{\epsilon^2} \log\left(\frac{1}{\epsilon}\right) \log\left(\frac{n}{\delta}\right)\right)$ |
| ST-Edge [14] | $O\left(\frac{m}{\epsilon^2} \log\left(\frac{m}{\delta}\right)\right)$ |
| MonteCarlo [34] | $O\left(\frac{m}{\epsilon^2} \log^4\left(\frac{1}{\epsilon}\right) \log\left(\frac{m}{\delta}\right)\right)$ |
| MonteCarlo-C [34] | $O\left(\frac{m}{\epsilon^2} \log^4\left(\frac{1}{\epsilon}\right) \log\left(\frac{m}{\delta}\right)\right)$ |
| Our TGT+ | $O\left(\frac{1}{\epsilon^2} \log^3\left(\frac{1}{\epsilon}\right) \log\left(\frac{m}{\delta}\right) \cdot \sum_{v_i \in \mathcal{V}} \frac{1}{d(v_i)} + m\right)$ |

The rationale of TGT+'s correctness has been explained in Section 4.1. For the time complexity, it comes from (i) the graph traversal in Lines 2-11, (ii) the random walk in Lines 12-17, and (iii) accessing each neighbor of each node in Line 18 with a total time of $O(m)$. With the adaptive switch condition in Eq.(14), TGT+ ensures that the cost of the first part does not exceed the second part, resulting in the cost of both is $O\left(\sum_{v_i \in \mathcal{V}} \sum_{v_j \in \mathcal{N}(v_i)} 2 \cdot \tau_{i,j} \cdot R(e_{i,j}, \tau_{i,j})\right)$, where $R(e_{i,j}, \tau_{i,j}) = O\left(\frac{\tau_{i,j}^2 \cdot \log\left(\frac{m}{\delta}\right)}{d(v_i)^2 \cdot \epsilon^2}\right)$ as Eq. (13) and $\tau_{i,j} = O\left(\log\left(\frac{1}{\epsilon}\right)\right)$ as Line 1 of Algorithm 1. Hence, the time complexity of TGT+ turns to the formula in Theorem 4.4. Table 2 compares the expected time of the randomized algorithm for $\epsilon$-approximate AESC computation. Notably, TGT+ eliminates an $m$ term in its bound, where the term $\sum_{v_i \in \mathcal{V}} \frac{1}{d(v_i)}$ can be simplified as $O(n)$ or even $O(n/\log n)$ using Kantorovich inequality on *scale-free* graphs with $m/n = O(\log n)$, manifesting the superiority of TGT+ over existing solutions.

## 5 EXPERIMENTS

In this section, we introduce the experimental settings, followed by evaluating our truncation bound and showing the performance of the proposed TGT+. At last, we analyze the sensitivity of constants $\gamma$ and $\omega$ in TGT+. All experiments are conducted on a Linux machine with Intel Xeon(R) Gold 6240@2.60GHz CPU and 377GB RAM in single-thread mode. None of the experiments need anywhere near all the memory. Due to space limitations, we refer interested readers to Appendix A for the scalability test.

### 5.1 Experimental Setups

**Datasets and groundtruths.** We include 5 different types of real undirected graphs at different scales, whose statistics are shown in Table 3. All datasets are collected from SNAP [21] and used as datasets in previous works [14, 29, 34]. For each graph, we generate groundtruth AESC by first computing $\mathbf{P}^\tau$ with $0 \le \tau \le 1000$ in parallel and then assembling them into SC by Eq.(2).

**Methods and parameters.** We compare TGT and TGT+ with three recent algorithms for AESC: ST-Edge [14], MonteCarlo [34] and MonteCarlo-C [34], as introduced in Section 2.3. We exclude Fast-Tree [29] from the competitors since it mainly offers relative approximation guarantees and is empirically shown significantly inferior to ST-Edge in [14]. Amid them, MonteCarlo and MonteCarlo-C are adapted for AESC computation, and the detailed modification is explained in Section 5.2. For the randomized algorithms TGT+, ST-Edge, MonteCarlo, and MonteCarlo-C, we follow [14] and set failure probability $\delta = 1/n$. Regarding MonteCarlo-C, we adopt the heuristic settings of $\beta_i$ as suggested in [34], since they are unknown. For the proposed TGT and TGT+, we set the constants $\gamma = 10$ and $\omega = 128$, unless otherwise specified. For a fair comparison, all tested algorithms are implemented in C++ and compiled by g++ 7.5 with −O3 optimization. For reproducibility, the source code is available at: https://github.com/jeremyzhangsq/AESC.

Table 3: Datasets.

| Name | #nodes | #edges |
|---|---|---|
| Facebook [30] | 4,039 | 88,234 |
| HepPh [20] | 34,401 | 420,784 |
| Slashdot [22] | 77,360 | 469,180 |
| Twitch [35] | 168,114 | 6,797,557 |
| Orkut [50] | 3,072,441 | 117,185,082 |

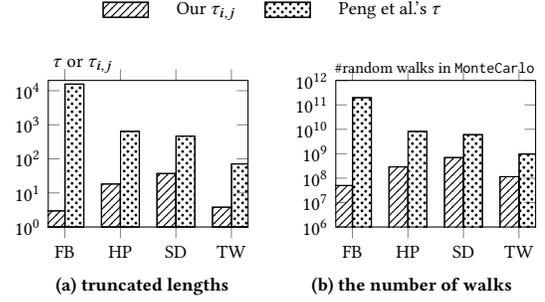

Figure 1: Our $\tau_{i,j}$ vs. Peng et al.'s $\tau$.

(a) truncated lengths  (b) the number of walks

### 5.2 Empirical Study of $\tau_{i,j}$ and $\tau$

In the first set of experiments, we evaluate the performance of the proposed truncated length in Section 3.1. Figure 1(a) reports the average of each edge $e_{i,j}$'s $\tau_{i,j}$ derived from Algorithm 1 and Peng et al.'s $\tau$ computed by Eq. (3) when $\epsilon = 0.01$ on Facebook (FB), HepPh (HP), Slashdot (SD) and Twitch (TW). We observe that our $\tau_{i,j}$ can significantly improve Peng et al.'s $\tau$ on all tested graphs. Notably, our $\tau_{i,j}$ is up to 3 orders of magnitude better than Peng et al.'s $\tau$. Correspondingly, the computational overhead of MonteCarlo can be reduced by replacing Peng et al.'s $\tau$ with our $\tau_{i,j}$. We take MonteCarlo as an example to demonstrate its superiority. Figure 1(b) reports the average number of random walks, where the major overhead of MonteCarlo stems from, for estimating each SC. Akin to the observation from Figure 1(a), MonteCarlo with our truncated lengths $\tau_{i,j}$ requires at most 3 orders of magnitude fewer random walks than Peng et al.'s $\tau$.

It is worth noting that MonteCarlo and MonteCarlo-C are designed for computing SC of a single node pair. Although our $\tau_{i,j}$ can remarkably cut down the number of random walks for an edge, there remain redundant random walks if invoking them for all edges individually. Hence, we further adapt MonteCarlo and MonteCarlo-C for efficient AESC computation by following the idea in TGT and TGT+ that iterate over each node. To summarize, for each node $v_i$, the adapted MonteCarlo and MonteCarlo-C first compute the largest $\tau_p$ among $v_i$'s local neighborhood as Line 2 in Algorithm 2, and then compute the number of samplings based on $\tau_p$. In the end, these extensions generate the corresponding random walks from $v_i$ and estimate $s_\tau(e_{i,j})$ for each $v_j \in \mathcal{N}(v_i)$.

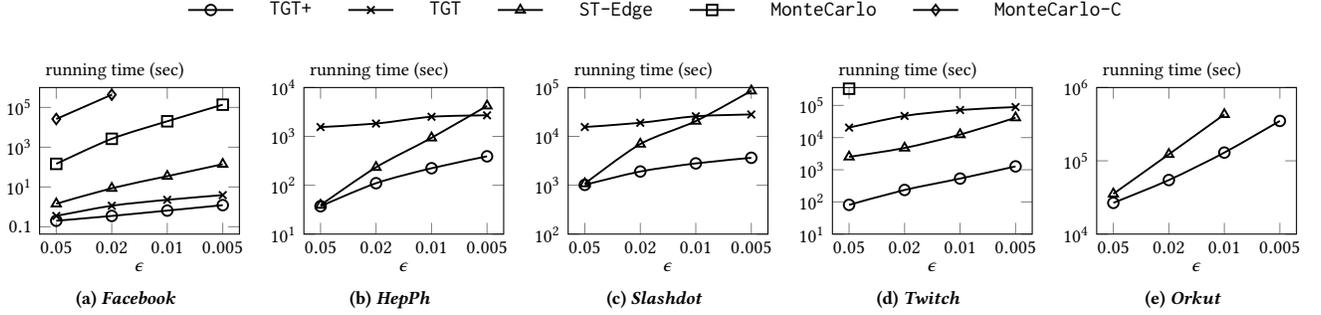

Figure 2: Running time of each algorithm by varying $\epsilon$.

## 5.3 Performance Evaluation

In the second set of experiments, we evaluate the performance of each approach in terms of efficiency and accuracy. For efficiency, we report the average running times (measured in wall-clock time) after all input data are loaded into the memory. For accuracy, we measure the actual average absolute error of the estimated all edge SC returned by each algorithm on each dataset. We run each algorithm with various $\epsilon$ in $\{0.05, 0.02, 0.01, 0.005\}$, and report the average evaluation score after repeating 3 trials. A method is excluded if it fails to report the result within 120 hours.

*5.3.1 Running time.* We first compare TGT+ with TGT and other competitors in terms of efficiency. Figure 2 reports each solution's running time for solving AESC with various $\epsilon$ settings. Benefiting from the truncation bound and the seamless combination of TGT and random walk samplings, the proposed TGT+ outperforms all competitors on all tested graphs and $\epsilon$ settings. Most notably, TGT+ improves the best competitor ST-Edge by at least one order of magnitude on Facebook and Twitch. We find that the improvement achieved by TGT+ becomes more remarkable as $\epsilon$ decreases. For example, TGT+ is 10.8× (resp. 23.5×) faster than ST-Edge on HepPh (resp. Slashdot) when $\epsilon = 0.005$. In addition, on the Orkut graph with 117 million edges, TGT+ is the only algorithm that can finish under all $\epsilon$ settings, demonstrating the scalability of our algorithm.

To evaluate the performance of the combination in TGT+, we next compare TGT+, TGT, MonteCarlo, and MonteCarlo-C, as all of them employ the edge-wise $\tau$ for the sake of fairness. As shown in Figure 2, MonteCarlo and MonteCarlo-C fail to return results within the allowed time limit in most cases. In particular, MonteCarlo can only terminate within 120 hours on Facebook and on Twitch when $\epsilon = 0.05$. The running time of MonteCarlo-C is even worse and is only feasible on Facebook with $\epsilon = 0.02, 0.05$. In contrast, TGT speeds up MonteCarlo and MonteCarlo-C by at least 2 orders of magnitude, demonstrating the superiority of the graph traversal in Section 3.2. However, TGT is still rather costly in comparison to TGT+. Specifically, TGT is only comparable to TGT+ on Facebook and is inferior to TGT+ on the rest graphs. For instance, TGT costs at least 1 and 2 orders of magnitude more time than TGT+ on Slashdot and Twitch, respectively, demonstrating the effectiveness of integrating deterministic traversal with randomized simulations in TGT+. To explain, the truncated length $\tau_{i,j}$ on the rest graphs (e.g., HepPh and Slashdot) is longer than that on Facebook, substantially increasing the overhead incurred by the graph traversal.

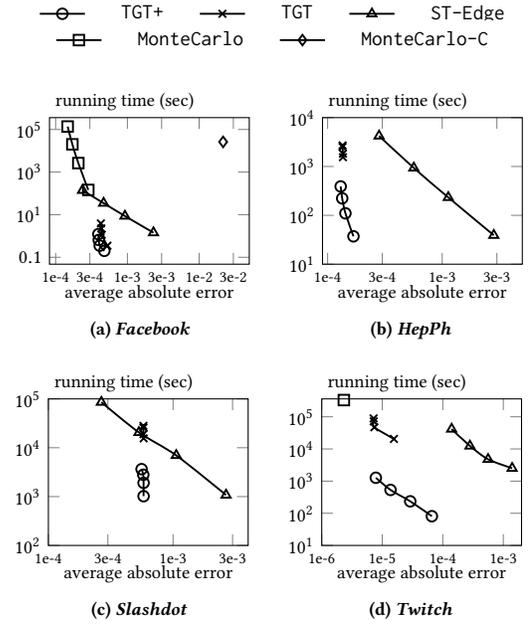

Figure 3: Tradeoffs between running time and absolute error.

*5.3.2 Accuracy.* We next report the tradeoffs between average absolute error (in $x$-axis) and running time (in $y$-axis) in Figure 3. The results are sorted in the ascending order of $\epsilon$, and the error-time curve closer to the lower left corner indicates a better performance. As shown, the overall observation is that TGT+ outperforms all competitors by achieving lower errors with less running time on all graphs. In particular, TGT+ achieves an average absolute error of 1.37E-05 with a time of 532 seconds on Twitch, while the closest solution TGT achieves an average absolute error of 1.56E-05 using over 20,000 seconds ($\approx 5.6$ hours). Regarding TGT, we observe that, under the same $\epsilon$ setting, the actual absolute error of TGT is slightly smaller than TGT+. This is as expected since TGT leverages the largest $\tau_p = \max_{v_j \in \mathcal{N}(v_i)} \tau_{i,j}$ as the maximal iteration for $v_i$. In other words, the SC value for the edge $e_{i,j}$ is overestimated if $\tau_{i,j} < \tau_p$. Furthermore, we notice that the absolute error of MonteCarlo-C is an order of magnitude larger than the closest competitor ST-Edge on Facebook. This is due to that the heuristic settings [34] for input parameters $\beta_i$ do not ensure the returned values are $\epsilon$-approximate.

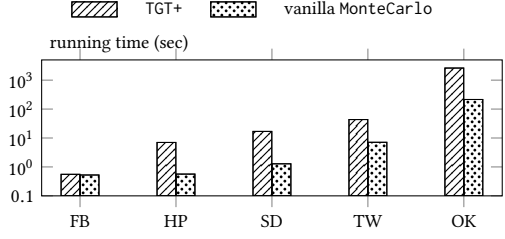

Figure 4: Preprocessing time of TGT+ and vanilla MonteCarlo.

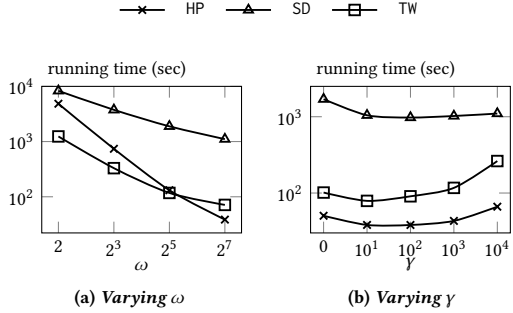

Figure 5: Varying constants in TGT+.

*5.3.3 Preprocessing time.* Recall that TGT, TGT+, MonteCarlo, and MonteCarlo-C rely on the eigen decomposition of the matrices pertaining to $\mathcal{G}$, e.g., $\mathbf{P}$ and $\mathbf{D}^{\frac{1}{2}}\mathbf{P}\mathbf{D}^{-\frac{1}{2}}$, in the preprocessing stage. In particular, MonteCarlo and MonteCarlo-C require the second largest eigenvalues, while TGT and TGT+ need the $\omega = 128$ largest eigenvalues and eigenvectors. Fortunately, by virtue of well-established techniques [33] and tools [19] for large-scale eigen decomposition, we can quickly obtain the desired eigenvalues and eigenvectors. Figure 4 reports the preprocessing time for TGT+ and vanilla MonteCarlo. As expected, the preprocessing time of TGT+ is comparable to MonteCarlo. In addition, compared to the running time for AESC displayed in Figure 2, the preprocessing costs are insignificant. For instance, the preprocessing time of TGT+ is about 45 minutes for the Orkut (OK) graph with 117 million edges, whereas the running time is at least 7 hours. Notice that this preprocessing step only needs to be conducted once for a graph.

## 5.4 Parameter Analysis

In the last set of experiments, we study the effects of TGT+'s constant: (i) $\omega$, the number of largest eigenvalues and eigenvectors of $\mathbf{D}^{\frac{1}{2}}\mathbf{P}\mathbf{D}^{-\frac{1}{2}}$ in Algorithm 1; (ii) $\gamma$, the number of candidates in Algorithm 3. In the sequel, we set $\epsilon = 0.05$ unless otherwise specified.

*5.4.1 Varying $\omega$.* Figure 5(a) reports the running time of TGT+ by setting $\gamma = 10$ and varying $\omega \in \{2, 8, 32, 128\}$ on HepPh (HP), Slashdot (SD) and Twitch (TW). As expected, TGT+ costs less running times as more eigenvalues and eigenvectors are exploited. Specifically, the improvement of $\omega$ is more remarkable on HepPh, where the running time of $\omega = 128$ is about 126× faster than $\omega = 2$. Besides, the running time of TGT+ achieves about 8× and 17× improvements by varying $\omega$ from 2 to 128 on SD and TW, respectively.

*5.4.2 Varying $\gamma$.* Figure 5(b) reports the running time of TGT+ by fixing $\omega = 128$ and picking $\gamma \in \{0, 10^1, 10^2, 10^3, 10^4\}$ for the computation of $\chi$ on HP, SD and TW. We observe that the running time of TGT+ first decreases and then increases as more candidates are considered. To explain, when $\gamma$ is too small, the upper bound $\chi$ for $|X|$ is too loose, rendering more random walks generated; when $\gamma$ is too large, Algorithm 3 incurs more computational overhead. For example, TGT+ with $\gamma = 0$ costs about 2× more time than that with $\gamma = 10$ on HP and SD. Meanwhile, TGT+ with $\gamma = 10,000$ costs over 3× more time than that with $\gamma = 10$ on TW.

## 6 ADDITIONAL RELATED WORK

In the sequel, we review existing studies germane to our work.

**Spanning centrality.** Apart from the methods discussed in Section 2.3, there exist several techniques for estimating SC (i.e., effective resistance (ER)). Fouss et al. [8] propose to calculate the exact ER values for all pairs of nodes in the input graph $G$ by first computing the Moore-Penrose pseudoinverse $\mathbf{L}^+$ of the Laplacian matrix $\mathbf{L} = \mathbf{D} - \mathbf{A}$, and then taking $\mathbf{L}^+[i, i] + \mathbf{L}^+[j, j] - \mathbf{L}^+[i, j] - \mathbf{L}^+[j, i]$ as the ER for any node pair $v_i, v_j \in V$. Teixeira et al. [40] and Mavroforakis et al. [29] utilize the random projection and symmetric diagonally dominant solvers to approximate the SC for all edges. After that, Jambulapati and Sidford [17] aime to compute the sketches of $\mathbf{L}$ and its pseudoinverse $\mathbf{L}^+$, and propose an algorithm for estimating ER values for all possible node pairs in $O(n^2/\epsilon)$ time. Besides MonteCarlo and MonteCarlo-C, Peng et al. [34] also propose two solutions by leveraging the connection between ER and the commute time [31] These works all focus on the $\epsilon$-multiplicative approximation and are beyond the scope of this paper.

**Personalized PageRank.** Another line of related work is personalized PageRank (PPR). In past decades, the efficient computation of PPR has been extensively studied in a plethora of works [2, 7, 16, 23–28, 38, 43–46, 49]. Among them, some recent approaches [16, 23, 24, 27, 28, 38, 43–46] also leveraged the idea of combining the deterministic graph traversal [2, 7, 26] with random walk simulations. At first glance, it seems that we can simply adapt and extend these techniques for computing $\epsilon$-approximate all edge SC. However, SC is much more sophisticated than PPR. This is due to that they are defined according to two inherently different types of random walks. More concretely, PPR leverages the one called *random walk with restart* (RWR) [41], which would stop at each visited node with a certain probability during the walk. In contrast, SC relies on simple random walks of various fixed lengths (from 1 to $\infty$), indicating that the walk in SC will not terminate as early as RWR does. Motivated by this, a linchpin of this work is a personalized truncation for the maximum random walk length. Correspondingly, the combination of graph traversal and random walks becomes more challenging.

## 7 CONCLUSION

In this paper, we propose two approximation algorithms for AESC computation. Our contributions consist of (i) enhanced lower bounds for truncating random walks, (ii) an algorithmic framework integrating the deterministic graph traversal with random walk sampling, and (iii) several carefully-designed optimization techniques for increasing efficiency. Our experiments on five real datasets demonstrate that our proposed algorithm significantly outperforms existing solutions in terms of practical efficiency without compromising theoretical and empirical accuracy. In the future, we plan to study AESC computation with relative error guarantees as well as under multithreading environments.

## A APPENDIX

### A.1 Proofs

**Proof of Theorem 3.2.** By Eq. (4) in Lemma 3.1 and the fact $\vec{f}_1 = 1$,

$$\frac{p_\ell(v_i, v_i)}{d(v_i)} + \frac{p_\ell(v_j, v_j)}{d(v_j)} - \frac{2p_\ell(v_i, v_j)}{d(v_j)} = \sum_{k=2}^{n} (\vec{f}_k[i] - \vec{f}_k[j])^2 \cdot \frac{\lambda_k^\ell}{2m}. \tag{18}$$

Consider $\ell = 0$. From Eq. (18), we have

$$\frac{p_0(v_i, v_i)}{d(v_i)} + \frac{p_0(v_j, v_j)}{d(v_j)} - \frac{2p_0(v_i, v_j)}{d(v_j)}$$
$$= \frac{1}{2m} \sum_{k=2}^{n} (\vec{f}_k[i] - \vec{f}_k[j])^2 = \frac{1}{d(v_i)} + \frac{1}{d(v_j)}.$$

Consider $\ell = 1$. Since $e_{i,j} \in \mathcal{E}$, we have
$$\frac{p_1(v_i, v_i)}{d(v_i)} + \frac{p_1(v_j, v_j)}{d(v_j)} - \frac{2p_1(v_i, v_j)}{d(v_j)}$$
$$= \frac{1}{2m} \sum_{k=2}^{n} (\vec{f}_k[i] - \vec{f}_k[j])^2 \cdot \lambda_k = \frac{-2}{d(v_i) \cdot d(v_j)}.$$

Combining these two equations yields
$$\sum_{k=2}^{n} (\vec{f}_k[i] - \vec{f}_k[j])^2 \cdot \frac{1+\lambda_k}{2m} = \frac{1}{d(v_i)} + \frac{1}{d(v_j)} - \frac{2}{d(v_i) \cdot d(v_j)} \quad (19)$$

Note that $1 = |\lambda_1| > |\lambda_2| > \cdots > |\lambda_n| > 0$. For simplicity, we let $\tau = \tau_{i,j}$ here. With Eq. (18) and Eq. (19), suppose $\tau$ and $t$ are odd numbers and $t \leq \tau$, we obtain
$$|s(e_{i,j}) - s_\tau(e_{i,j})| = \left|\sum_{\ell=\tau+1}^{\infty} \frac{p_\ell(v_i, v_i)}{d(v_i)} + \frac{p_\ell(v_j, v_j)}{d(v_j)} - \frac{2p_\ell(v_i, v_j)}{d(v_j)}\right|$$
$$= \left|\frac{1}{2m} \sum_{k=2}^{n} (\vec{f}_k[i] - \vec{f}_k[j])^2 \sum_{\ell=\tau+1}^{\infty} \lambda_k^\ell\right|$$
$$= \frac{1}{2m} \sum_{k=2}^{n} (\vec{f}_k[i] - \vec{f}_k[j])^2 \cdot \frac{\lambda_k^{\tau+1}}{1-\lambda_k}$$
$$= \frac{1}{2m} \sum_{k=2}^{n} (\vec{f}_k[i] - \vec{f}_k[j])^2 \cdot (1+\lambda_k) \cdot \frac{\lambda_k^{\tau+1}}{1-\lambda_k^2}$$
$$= \Delta_\tau + \frac{1}{2m} \sum_{k=\omega}^{n} (\vec{f}_k[i] - \vec{f}_k[j])^2 \cdot (1+\lambda_k) \cdot \frac{\lambda_k^{\tau+1}}{1-\lambda_k^2}$$
$$\leq \Delta_t + \frac{1}{2m} \sum_{k=\omega}^{n} (\vec{f}_k[i] - \vec{f}_k[j])^2 \cdot (1+\lambda_k) \cdot \frac{\lambda_k^{\tau+1}}{1-\lambda_k^2}$$
$$\leq \Delta_t + \frac{\lambda_\omega^{\tau+1}}{1-\lambda_\omega^2} \cdot \left(\frac{1}{2m} \sum_{k=2}^{n} (\vec{f}_k[i] - \vec{f}_k[j])^2 \cdot (1+\lambda_k) - \Upsilon\right)$$
$$= \Delta_t + \frac{\lambda_\omega^{\tau+1}}{1-\lambda_\omega^2} \cdot \left(\frac{1}{d(v_i)} + \frac{1}{d(v_j)} - \frac{2}{d(v_i) \cdot d(v_j)} - \Upsilon\right).$$

Plugging Eq. (5) into the above inequality yields $|s(e_{i,j}) - s_\tau(e_{i,j})| \leq \frac{\epsilon}{2}$, which proves the theorem.

**Proof of Theorem 3.3.** In the $\ell$-th iteration of the source $v_i \in \mathcal{V}$, the graph traversal operation (Lines 6-9) is equivalent to the sparse matrix-vector multiplication $p_\ell(\cdot, v_i) = \mathbf{P} \cdot p_{\ell-1}(\cdot, v_i)$. Note that $\tau_p = \max_{v_j \in \mathcal{N}(v_i)} \tau_{i,j}$, where $\tau_{i,j}$ ensures $|s(e_{i,j}) - s_\tau(e_{i,j})| \leq \epsilon$ in terms of Theorem 3.2. Correspondingly, when $\tau_p$ iterations terminate for each $v_i \in \mathcal{V}$, the truncated $s_\tau(e_{i,j})$ computed by Eq. (2) is $\epsilon$-approximate. The worst-case complexity of Algorithm 2 is $O(n \cdot m \cdot \tau_p)$, since $\tau_p$ steps of graph traversals are conducted from all $n$ nodes and each invocation of graph traversal costs $O(m)$. For any $v_i \in \mathcal{V}$, as Line 1 of Algorithm 1,
$$\tau_p \leq \log_{\frac{1}{|\lambda_2|}} \left(\frac{\frac{1}{d(v_i)} + \frac{1}{d(v_j)} - \frac{2}{d(v_i) \cdot d(v_j)}}{\frac{\epsilon}{2} \cdot (1-|\lambda_2|^2)}\right) = O\left(\log(\frac{1}{\epsilon})\right),$$

the worst-case complexity turns to $O\left(n \cdot m \cdot \log(\frac{1}{\epsilon})\right)$, which completes the proof.

**Proof of Lemma 4.1.** Based on Eq.(9), $g_\tau(v_i, v_j)$ turns to $g_\tau(v_i, v_j) = h_\tau(v_i, v_i) - h_\tau(v_i, v_j) = h_{\tilde{\tau}}(v_i, v_i) - h_{\tilde{\tau}}(v_i, v_j) + g_{\tilde{\tau}\to\tau}(v_i, v_j)$. Note that
$$g_{\tilde{\tau}\to\tau}(v_i, v_j) = \sum_{\ell=\tilde{\tau}+1}^{\tau} \frac{p_\ell(v_i, v_i)}{d(v_i)} - \frac{p_\ell(v_i, v_j)}{d(v_j)}$$
$$= \sum_{\ell=\tilde{\tau}+1}^{\tau} \sum_{v_x \in \mathcal{V}} p_{\tilde{\tau}}(v_i, v_x) \cdot \frac{p_{\ell-\tilde{\tau}}(v_x, v_i)}{d(v_i)} - p_{\tilde{\tau}}(v_i, v_x) \cdot \frac{p_{\ell-\tilde{\tau}}(v_x, v_j)}{d(v_j)}$$

By the fact that $\frac{p_{\ell-\tilde{\tau}}(v_x, v_i)}{d(v_i)} = \frac{p_{\ell-\tilde{\tau}}(v_i, v_x)}{d(v_x)}$ and $\frac{p_{\ell-\tilde{\tau}}(v_x, v_j)}{d(v_j)} = \frac{p_{\ell-\tilde{\tau}}(v_j, v_x)}{d(v_x)}$, the above equation becomes
$$\sum_{\ell=\tilde{\tau}+1}^{\tau} \frac{p_\ell(v_i, v_i)}{d(v_i)} - \frac{p_\ell(v_i, v_j)}{d(v_j)}$$
$$= \sum_{v_x \in \mathcal{V}} \frac{p_{\tilde{\tau}}(v_i, v_x)}{d(v_x)} \cdot \left(\sum_{\ell=1}^{\tau_j - \tilde{\tau}} p_\ell(v_i, v_x) - p_\ell(v_j, v_x)\right)$$
$$= \frac{1}{d(v_i)} \sum_{v_x \in \mathcal{V}} p_{\tilde{\tau}}(v_x, v_i) \cdot \left(\sum_{\ell=1}^{\tau_j - \tilde{\tau}} p_\ell(v_i, v_x) - p_\ell(v_j, v_x)\right),$$

which completes the proof.

**Proof of Lemma 4.3.** Let $v_{w_j}$ be the $j$-th node on $W_i$. Note that for any two adjacency nodes $v_{w_j}, v_{w_{j+1}}$ on $W_i$, $(v_{w_j}, v_{w_{j+1}})$ is an edge in $\mathcal{G}$.
$$\sum_{v_x \in W_i} p_{\tilde{\tau}}(v_x, v_i)$$
$$= \frac{1}{2} p_{\tilde{\tau}}(v_{w_1}, v_i) + \frac{1}{2} p_{\tilde{\tau}}(v_{w_\ell}, v_i) + \frac{1}{2} \sum_{j=1}^{\ell-1} p_{\tilde{\tau}}(v_{w_j}, v_i) + p_{\tilde{\tau}}(v_{w_{j+1}}, v_i)$$
$$\leq \frac{1}{2} p_{\tilde{\tau}}(v_{w_1}, v_i) + \frac{1}{2} p_{\tilde{\tau}}(v_{w_\ell}, v_i)$$
$$\quad + \frac{\ell-1}{2} \cdot \max_{1 \leq j \leq \ell-1} \{p_{\tilde{\tau}}(v_{w_j}, v_i) + p_{\tilde{\tau}}(v_{w_{j+1}}, v_i)\}$$
$$\leq \frac{1}{2} \max_{v_l \in \mathcal{N}(v_i)} \{p_{\tilde{\tau}}(v_l, v_i)\} + \frac{1}{2} \max_{v_l \in \mathcal{V}} \{p_{\tilde{\tau}}(v_l, v_i)\}$$
$$\quad + \frac{\ell-1}{2} \cdot \max_{e_{x,y} \in \mathcal{E}} \{p_{\tilde{\tau}}(v_x, v_i) + p_{\tilde{\tau}}(v_y, v_i)\} = UB(v_i, \ell).$$

On the other hand,
$$\sum_{v_x \in W_i} p_{\tilde{\tau}}(v_x, v_i) \geq \min_{v_l \in \mathcal{N}(v_i)} \{p_{\tilde{\tau}}(v_l, v_i)\} + (\ell-1) \cdot \min_{v_l \in \mathcal{V}} \{p_{\tilde{\tau}}(v_l, v_i)\}$$
$$= LB(v_i, \ell).$$

The lemma is then proved.

**Proof of Theorem 4.4.** For a fixed edge $e_{i,j} \in \mathcal{E}$, denote estimation of $g_\tau(v_i, v_j)$ as $\hat{g}_\tau(v_i, v_j) = g_{\tilde{\tau}}(v_i, v_j) + X$, which is an unbiased estimator as mentioned in Section 4.1. As per Lemma 4.2 and Eq. (12), by setting $R(e_{i,j}, \tau_{i,j} - \tilde{\tau}) = \frac{8\chi^2 \log(\frac{2m}{\delta})}{d(v_i)^2 \cdot \epsilon^2}$, we obtain that

$$\mathbb{P}[|\hat{g}_\tau(v_i, v_j) - g_\tau(v_i, v_j)| \geq \epsilon/4]$$
$$\leq 2 \exp\left(-\frac{2 \cdot R(e_{i,j}, \tau_{i,j} - \tilde{\tau})^2 \cdot (\epsilon/4)^2}{R(e_{i,j}, \tau_{i,j} - \tilde{\tau}) \cdot (\chi/d(v_i))^2}\right)$$
$$= 2\exp\left(-\frac{\epsilon^2 \cdot d(v_i)^2}{8\chi^2} \cdot R(e_{i,j}, \tau_{i,j} - \tilde{\tau})\right)$$
$$= 2\exp\left(\log\frac{2m}{\delta}\right) = \frac{\delta}{m},$$

which further turns into

$$\mathbb{P}[|s(e_{i,j}) - \hat{s}(e_{i,j})| \geq \epsilon] \leq \frac{\delta}{m}$$

since Theorem 3.2 ensures that Line 2 of Algorithm 4 satisfying $|s(e_{i,j}) - s_\tau(e_{i,j})| \leq \epsilon/2$ and $s_\tau(e_{i,j}) = g_\tau(v_i, v_j) + g_\tau(v_j, v_i)$). Based on union bound, we can derive that Algorithm 4 returns $\epsilon$-approximate SC values $s_\tau(e_{i,j}) \, \forall e_{i,j} \in \mathcal{E}$ with the probability at least $1 - \delta$.

**Table 4: The running time of TGT+ by varying the number of edges.**

| #edges (×$10^6$) | 0.2 | 0.5 | 1 | 2 | 5 |
|---|---|---|---|---|---|
| time (seconds) | 0.357 | 0.921 | 2.400 | 5.163 | 12.106 |

**Table 5: The running time of TGT+ by varying the number of nodes.**

| #nodes (×$10^3$) | 2 | 5 | 10 | 20 | 50 |
|---|---|---|---|---|---|
| time (seconds) | 2.293 | 2.328 | 2.400 | 2.593 | 2.946 |

Regarding the time complexity, notice that, by Eq. (14), we ensure that the cost of the deterministic part does not exceed that of using random walk samplings. Hence, the overall time complexity of TGT+ is upper-bounded by

$$m + \sum_{v_i \in \mathcal{V}} \sum_{v_j \in \mathcal{N}(v_i)} \left(2 \cdot (\tau_{i,j} - \tilde{\tau}) \cdot R(e_{i,j}, \tau_{i,j} - \tilde{\tau}) + \sum_{\tau=0}^{\tilde{\tau}-1} R(e_{i,j}, \tau_{i,j} - \tau)\right) \quad (20)$$

$$\leq m + \sum_{v_i \in \mathcal{V}} \sum_{v_j \in \mathcal{N}(v_i)} 2 \cdot \tau_{i,j} \cdot R(e_{i,j}, \tau_{i,j})$$

As per Eq. (15), we can obtain that $R(e_{i,j}, \tau_{i,j}) \leq \frac{32 \cdot \tau_{i,j}^2 \cdot \log(\frac{2m}{\delta})}{d(v_i)^2 \cdot \epsilon^2}$. Correspondingly,

$$Eq.\ (20) \leq m + \sum_{v_i \in \mathcal{V}} \sum_{v_j \in \mathcal{N}(v_i)} \frac{64 \cdot \tau_{i,j}^3 \log(\frac{2m}{\delta})}{d(v_i)^2 \cdot \epsilon^2}.$$

Since $\tau_{i,j} \leq \log_{\frac{1}{|\lambda_2|}}\left(\frac{\frac{1}{d(v_i)} + \frac{1}{d(v_j)} - \frac{2}{d(v_i) \cdot d(v_j)}}{\frac{\epsilon}{2} \cdot (1 - |\lambda_2|^2)}\right) = O\left(\log(\frac{1}{\epsilon})\right)$ as Line 1 of Algorithm 1, we have

$$Eq.\ (20) = O\left(m + \frac{\log(\frac{m}{\delta})}{\epsilon^2} \sum_{v_i \in \mathcal{V}} \sum_{v_j \in \mathcal{N}(v_i)} \frac{\log^3(\frac{1}{\epsilon})}{d(v_i)^2}\right)$$
$$= O\left(m + \frac{\log(\frac{m}{\delta})}{\epsilon^2} \sum_{v_i \in \mathcal{V}} \frac{\log^3(\frac{1}{\epsilon})}{d(v_i)}\right)$$
$$= O\left(m + \frac{1}{\epsilon^2} \cdot \log^3(\frac{1}{\epsilon}) \cdot \log(\frac{m}{\delta}) \cdot \sum_{v_i \in \mathcal{V}} \frac{1}{d(v_i)}\right),$$

which completes the proof.

### A.2 Scalability Test

Besides the evaluation in Section 5, we also test the scalability of TGT+ on synthetic graphs of varying sizes generated by the Erdos Renyi random graph model. To evaluate scalability, we fix the number of nodes as $10^4$ (resp. the number of edges as $10^6$) and vary the number of edges from 0.2, 0.5, 1, 2, 5×$10^6$ (resp. the number of nodes from 2, 5, 10, 20, 50 × $10^3$). We have included the results in Table 4 and Table 5. Our results show that the running time grows linearly with the number of nodes and edges, confirming the time complexity of TGT+ and demonstrating its scalability.